\def\rn{\noindent\parshape 2 0truecm 8.8truecm 0.3truecm 8.5truecm}
\def\nn#1 #2{#1, #2.}				
\def\nnn#1 #2 #3{#1, #2. #3.}			
\def\nnnn#1 #2 #3 #4{#1, #2. #3. #4.}		
\def\nnnnn#1 #2 #3 #4 #5{#1, #2. #3. #4. #5.}	
\def\dualand{, \&\hbox{ }}				
\def\multiand{, \&\hbox{ }}				

\def\rg#1;#2;#3;#4;#5;#6 {\par\rn#1 #2, {\it #3}, {\bf #4}, #5 (``#6'') \par}
\def\rf#1;#2;#3;#4;#5 {\par\rn#1 #2, {\it #3}, {\bf #4}, #5\par}
\def\rfbook#1;#2;#3;#4;#5 {{\frenchspacing\par\rn#1 #2, {\it #3} (#4: #5)\par}}
\def\rfproc#1;#2;#3;#4;#5;#6 {{\frenchspacing\par\rn#1 #2, in {\it #3}, ed. #4 (#5: #6)\par}}
\def\rfprep#1;#2;#3  {{\par\rn#1 #2, #3\par}}
\def\rfprepp#1;#2;#3 {{\par\rn#1 #2, #3\par}}

\def\expec#1{\langle#1\rangle}
\def\bigexpec#1{\left\langle#1\right\rangle}

\def\etal{{\frenchspacing\it et al.}}
\def\ie{{\frenchspacing\it i.e.}}
\def\eg{{\frenchspacing\it e.g.}}

\def\rms{rms}

\def\beq#1{\begin{equation}\label{#1}}
\def\eeq{\end{equation}}
\def\beqa#1{\begin{eqnarray}\label{#1}}
\def\eeqa{\end{eqnarray}}
\def\eq#1{equation~(\ref{#1})}

\def\eqn#1{~(\ref{#1})}

\def\spose#1{\hbox to 0pt{#1\hss}}
\def\simlt{\mathrel{\spose{\lower 3pt\hbox{$\mathchar"218$}}
     \raise 2.0pt\hbox{$\mathchar"13C$}}}
\def\simgt{\mathrel{\spose{\lower 3pt\hbox{$\mathchar"218$}}
     \raise 2.0pt\hbox{$\mathchar"13E$}}}
\def\simpropto{\mathrel{\spose{\lower 3pt\hbox{$\mathchar"218$}}
     \raise 2.0pt\hbox{$\propto$}}}

\def\ed{\end{document}}

\font\bfmath=cmmib10
\def\vth{\hbox{\bfmath\char'002}}       
\def\vmu{\hbox{\bfmath\char'026}}	

\def\x{{\bf x}}
\def\C{{\bf C}}

\def\zbar{{\bar z}}
\def\dz{\Delta z}
\def\Om{\Omega_m}
\def\Ol{\Omega_\Lambda}
\def\dOm{\Delta\Om}
\def\dOl{\Delta\Ol}

\def\dm{\Delta m}
\def\F{{\bf F}}

\def\tr{\hbox{tr}\>}
\def\nth{n^{\rm th}}
\def\flum{d_{lum}}
\def\fang{d_{ang}}
\def\fage{d_{age}}
\def\fgr{d_{gr}}
\def\c{d}
\def\I{\eta}
\def\zrms{z_{rms}}
\def\zmax{{z_{max}}}
\def\nbar{{\bar{N}}}
\def\ignore#1{}

\documentstyle[emulateapj]{article}
\begin{document}

\journalid{337}{15 January 1989}
\articleid{11}{14}

\submitted{Submitted to ApJL May 11, 1998}

\title{COSMIC COMPLEMENTARITY:\\ 
PROBING THE ACCELERATION OF THE UNIVERSE
}

\author{
Max Tegmark$^{1,2}$, Daniel J. Eisenstein$^1$, Wayne Hu$^1$ and Richard G. Kron$^3$
}

\begin{abstract}
We assess the accuracy with which $\Om$ and $\Ol$ can be measured
by combining various types of upcoming experiments.
Useful expressions for the Fisher information matrix are derived
for classical cosmological tests involving luminosity (\eg, SN Ia), 
angular size, age and number counts. These 
geometric probes are found to be quite complementary both to each other
and to inferences from cluster abundance and the cosmic microwave background
(CMB).
For instance, a joint analysis of SN Ia and CMB 
reduces the error bars on $\Ol$ by about an order of magnitude compared to a
separate analysis of either data set.

\end{abstract}

\date{\today}

\makeatletter
\global\@specialpagefalse
\def\@oddfoot{
\ifnum\c@page>1
  \reset@font\rm\hfill \thepage\hfill
\fi
\ifnum\c@page=1
{\sl
Available in color from
h t t p://www.sns.ias.edu/$\tilde{~}$max/complementarity.html}
\hfill\\
\fi
} \let\@evenfoot\@oddfoot
\makeatother

\section{Introduction}

It may be possible to measure cosmological 
parameters with great accuracy using upcoming
cosmic microwave background (CMB) experiments
(Jungman {\etal} 1996; Bond {\etal} 1997; Zaldarriaga {\etal} 1997),
galaxy surveys
(Tegmark 1997; Goldberg \& Strauss 1998; Hu {\etal} 1998)
and supernova Ia (SN Ia) searches (Goobar \& Perlmutter 1995;
Perlmutter {\etal} 1998; Garnavich {\etal} 1998). 
However, no single type of measurement alone can constrain
all parameters, as it will inevitably suffer from
so-called {\it degeneracies} in which particular combinations of
changes in parameters leave the result essentially 
unaffected 
(Bond {\etal} 1994, 1997; 
Zaldarriaga {\etal} 1997; Metcalf \& Silk 1998;
Huey {\etal} 1998).
Fortunately, different types of cosmological measurements 
are often highly complementary, breaking each other's
degeneracies and combining to give much more accurate measurements
than any one could 
give 
alone. 
For example, CMB measurements are highly complementary to 
both galaxy surveys (Tegmark {\etal} 1997; Hu {\etal} 1998; 
Gawiser \& Silk 1998; Webster {\etal} 1998; Eisenstein {\etal} 1998)
and SN Ia 
(Zaldarriaga {\etal} 1997; Tegmark 1997; White 1997). 

The topic of this {\it Letter} is 
probes of the acceleration of the Universe,
given by the density parameters $\Om$ for matter and 
$\Ol$ for vacuum density (cosmological constant).
Most of the cosmological tests that we discuss are well-known. 
Our focus is on their degeneracy structure, {\ie}, on which ones
are complementary and which ones act as independent cross-checks 
of one another.
We address this by computing 
the {\it Fisher information matrix} $\F$ for each of the tests.
This has the advantage of explicitly
showing how the accuracy and degeneracy depends on the survey details.
It also allows a unified treatment of all tests, since 
if independent experiments are analyzed jointly, their
Fisher information matrices simply add.

\bigskip
\bigskip

{\footnotesize

$^1$ Institute for Advanced Study, Princeton, NJ 08540;\\ 
\hglue0.6cm max@ias.edu, eisenste@ias.edu, whu@ias.edu
 
\vskip0.1cm
     
$^2$ Hubble Fellow

$^3$ FNAL, MS 127, Batavia, IL 60510
}

\bigskip
\goodbreak

\section{Calculation of the Fisher matrices}

All data sets discussed below consist of a vector $\x$ of
measured numbers $x_1,...,x_N$ whose
probability distribution  $f(\x;\vth)$ depends on a vector of cosmological parameters 
$\vth$ that we wish to estimate. 
In our case, $\theta_1=\Om$ and $\theta_2=\Ol$.
The {\it Fisher information matrix} for a data set
(see Tegmark {\etal} 1997 for a comprehensive review), defined as
\beq{FisherDefEq}
\F_{ij} \equiv - \bigexpec{{\partial^2 \ln f\over\partial\theta_i\partial\theta_j}},
\eeq
quantifies its information content about these parameters.
Its inverse $\F^{-1}$ gives the best attainable covariance 
matrix for the measurement errors on these parameters,
illustrated by the error ellipses in Fig.~2.
We will now specify probability distributions $f$ for
the various cosmological tests and compute the 
corresponding Fisher matrices.

\subsection{Luminosity, size, age and clustering}

The cosmological tests based on luminosity, angular size, age and clustering
(see {\eg} Weinberg 1972, hereafter W72; Peebles 1993),
can all be described as noisy measurements of some quantities
$x_n$ at redshifts $z_n$, $n=1,...,N$. We model them as
\beq{xEq}
x_n = a\ln d(z;\Om,\Ol) + b + \varepsilon_n,
\eeq
where $a$ and $b$ are constants independent of 
$\Om$ and $\Ol$, the function $d$ incorporates the effects of cosmology 
and $\varepsilon_n$ is a random term with zero
mean ($\expec{\varepsilon_n}=0$)
including all sources of measurement error.

For luminosity tests like SN Ia, $x_n$ is the observed magnitude 
of the $\nth$ object and $d$ is the luminosity distance (W72):
\beq{LumDistEq}
\flum  = (1+z){S(\kappa\I)\over\kappa},\quad
\I(z;\Om,\Ol)
=\int_0^z{dz'\over E(z')},
\eeq
\beq{Aeq}
E(z)\equiv\left[(1+z)^2(1+\Om z)-z(2+z)\Ol\right]^{1/2},
\eeq
where $\kappa\equiv\sqrt{|1-\Om-\Ol|}$.
We recognize $E=H(z)/H_0$ as the relative expansion rate at an
earlier time and $1/H_0\kappa$ 
as (the magnitude of) the current radius of curvature of the Universe. From 
the definition of magnitudes, $a=5/\ln 10$.
The errors $\varepsilon_n$ include 
errors in extinction correction and intrinsic scatter in the ``standard candle''
luminosity.

For tests involving the observed angular sizes $\theta_n$ of objects
at redshifts $z_1,...,z_N$, we define $x_n\equiv\ln\theta_n$, $a=-1$ and 
take $d$ to be the the angular size distance (W72): $\fang=\flum/(1+z)^2$.
For such tests (see {\eg} Daly 1998; Pen 1997), $\varepsilon_n$
includes scatter in the ``standard yardstick'' size.

For tests involving estimates $t_n$ of the age of the Universe
at redshifts $z_n$, we define $x_n\equiv\ln H_0 t_n$. Setting $a=1$, this
gives (W72)
\beq{AgeEq}
\fage =\int_z^\infty{dz'\over(1+z')E(z')}.
\eeq

For tests involving the observed growth $G_n$ in the amplitude of linear density 
fluctuations since redshift $z_n$,
we choose $x_n\equiv-\ln G_n$, $a=1$ and take $\c$ to be the linear growth factor (W72):
\beq{GrowthEq}
\fgr\equiv {D(z)\over D(0)},\quad D(z)\propto E(z)\int_z^\infty{(1+z)\over E(z)^3}dz.
\eeq

Assuming that the errors $\varepsilon_n$ have a Gaussian distribution,
the Fisher matrix is given by (Tegmark {\etal} 1997) 
\beq{GaussianFeq}
\F_{ij} = {1\over 2}\tr[\C^{-1}\C,_i\C^{-1}\C,_j] + \vmu,_i^t\C^{-1}\vmu,_j,
\eeq
where $\vmu\equiv\expec{\x}$ is the mean $[\mu_n=\ln\c(z_n)]$ and
$\C\equiv\expec{\x\x^t}-\vmu\vmu^t$ is the covariance matrix of $\x$.
Commas denote derivatives, so 
$\vmu,_i\equiv\partial\vmu/\partial\theta_i$.
For simplicity, we will assume that 
\beq{Ceq}
\C_{mn}=\delta_{mn}\sigma_n^2,
\eeq
\ie, that all the magnitude errors 
$\varepsilon_n$ are uncorrelated.
Our treatment below is readily generalized to 
non-diagonal error 
models $\C$, more appropriate for describing systematics.
Since $\C,_i=0$, all the information 
about $\Om$ and $\Ol$ comes from the second term in \eq{GaussianFeq},
giving
\beq{Feq2}
\F_{ij} = a^2\sum_{n=1}^N  
{1\over\sigma_i^2}
{\partial\ln\c\over\partial\theta_i}(z_n)
{\partial\ln\c\over\partial\theta_j}(z_n).
\eeq

\subsection{A supernova example}

To bring out the physics, let us evaluate this explicitly for the 
SN Ia example --- the other cases are analogous. 
SN Ia have had their accuracy assessed previously,
first by Goobar \& Perlmutter (1995)
and subsequently by making $\chi^2$-fits to real data 
(Perlmutter {\etal} 1998; Garnavich {\etal} 1998; White 1998);
however, this is the first treatment involving their Fisher matrix.

In this illustration, we take all magnitude errors to be 
equal, $\sigma_i=\dm$.  It is instructive to rewrite \eq{Feq2} as
\beq{Feq3}
\F_{ij} = {N\over(\dm)^2}\int_0^\infty g(z) w_i(z) w_j(z) dz,
\eeq
where
\beq{wEq}
w_i(z) \equiv 
{5\over\ln 10}
\left\{{\kappa S'[\kappa\I(z)]\over S[\kappa\I(z)]}
\left[{\partial\I\over\partial\theta_i} - {\I(z)\over  2\kappa^2}\right]
+ {1\over  2\kappa^2}
\right\},
\eeq
\beqa{PartialEq}
{\partial\I\over\partial\Om}(z) &=& -{1\over 2}\int_0^z{z'(1+z')^2\over E(z')^3}dz',\\
{\partial\I\over\partial\Ol}(z) &=&  {1\over 2}\int_0^z{z'(2+z')\over E(z')^3}dz',
\eeqa
and the SN Ia redshift distribution is given by
$g(z) = {1\over N}\sum_{n=1}^N \delta(z-z_n)$.
The expression in braces approaches 

\smallskip
\smallskip

\centerline{{\vbox{\epsfxsize=9.0cm\epsfbox{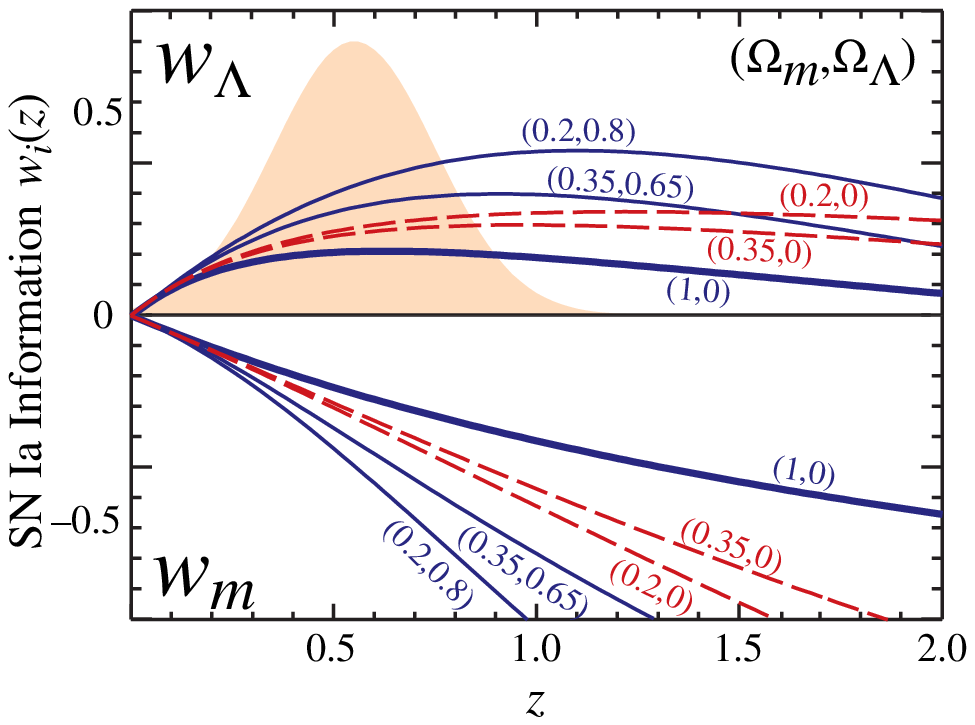}}}}
{\footnotesize {\bf FIG. 1}
--- The SN Ia weight functions $w_\Lambda$ (positive)
and $w_m$ (negative) are plotted for standard CDM,
two open $(\Ol=0)$ models and two flat $(\Ol=1-\Om)$ models. 
The Fisher matrix element $\F_{ij}$ is  
computed by simply integrating the product of the curves $w_i$ and $w_j$ and a
redshift distribution $f$ such as the shaded one.
}

\bigskip

\centerline{{\vbox{\epsfxsize=9.0cm\epsfbox{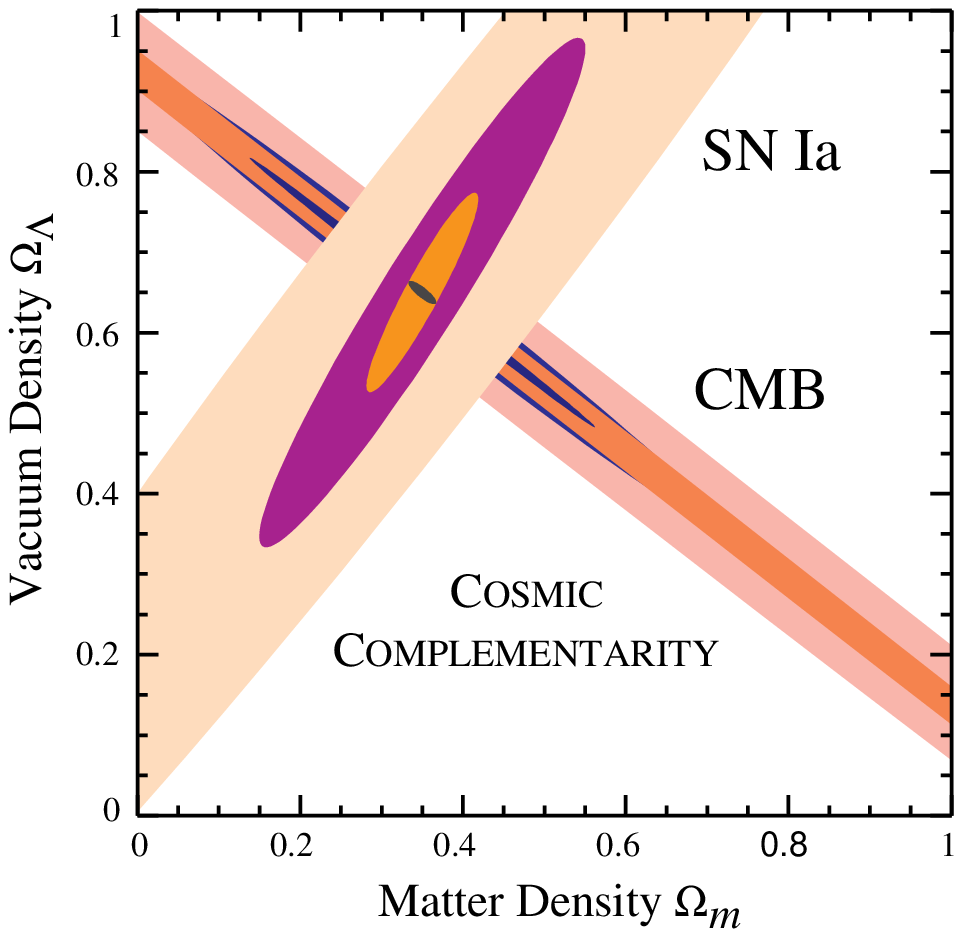}}}}
{\footnotesize {\bf FIG. 2}
--- $68\%$ confidence regions are shown for 
the upcoming CMB experiments and hypothetical SN Ia data
sets specified in Table 1.
The assumed fiducial model is 
COBE-normalized $\Lambda$CDM with $\Om=0.35$, $\Ol=0.65$, $\Omega_b=0.05$, 
and $h=0.65$.
Combining the CMB and SN Ia data shrinks the error region to the 
overlap of the two corresponding ellipses: 
for instance, a joint analysis of the optimistic SN Ia case with 
polarized Planck data gives the tiny black ellipse in the center.
}

\smallskip
\goodbreak

\noindent
$\I^{-1} \partial\I/\partial\theta_i -\I^2/6$ as $\kappa\to 0$.
The contribution to $\F$ from each redshift can thus be split
into two factors, one
reflecting the quality of the data set ($N g[z]/\dm^2$)
and the other incorporating the effects of cosmology
(the weight functions $w_i$).
The functions $w_i$ are plotted in Figure 1 for a variety of
cosmological models.  

If all the observed supernovae were at the same
redshift $z$, then the resulting $2\times 2$ Fisher matrix
$\F_{ij}\propto w_i(z) w_j(z)$ would have rank 1, \ie, be singular.
The vanishing eigenvalue would correspond to the eigenvector
$(w_\Omega,-w_\Lambda)$.  Physically, this is because there is more
than one way of fitting a single measured quantity $\flum(z)$ by varying
two parameters ($\Om$ and $\Ol$).  The corresponding ellipse in Figure
2 would be infinitely long, with slope $-w_\Omega/w_\Lambda$, the
ratio of the magnitudes of the $\Om$ and $\Ol$ curves in Figure 1 at
that redshift.  The SN Ia ellipses plotted in Figure 2 correspond to a
range of redshifts, with $f$ being a Gaussian of mean $\zbar$ and
standard deviation $\dz$ given by Table~1. This breaks the degeneracy
only marginally, leaving the SN ellipses quite skinny, since the ratios
$w_\Omega/w_\Lambda$ in Figure 1 are seen to vary only weakly with $z$. 

\subsection{Counts}

For a sample of objects volume limited out to redshift $\zmax$,
the average number per unit redshift is (W72)
\beq{pEq}
p(z)\propto{\flum(z)^2\over (1+z)^2 E(z)}.
\eeq
Defining $x_i=z_i$, the probability distribution for the observed set
of $N$ redshifts $\x$ is not a multivariate Gaussian as above, but a
multivariate Poisson distribution,
\beq{PoissonEq}
f(\x) = e^{-\nbar}{\nbar^N\over N!}\prod_{n=1}^N g(z_n),
\eeq
where $\nbar\equiv\expec{N}=\int_0^\zmax p(z)dz$ is the expected number of objects and
$g(z)\equiv p(z)/\nbar$ can be interpreted as a probability distribution for
the redshift of a typical object. Note that the integer $N$ is itself random, with a 
Poisson distribution. 
Substituting equations\eqn{pEq} and\eqn{PoissonEq} into\eqn{FisherDefEq} gives
\beq{CountFisherEq}
\F_{ij} = -\nbar\int_0^\zmax {\partial^2\ln g\over\partial\theta_i\partial\theta_j}g(z)dz 
+ {1\over\nbar}
{\partial\nbar\over\partial\theta_i}{\partial\nbar\over\partial\theta_j}.
\eeq
We will neglect the last term to be conservative, since it reflects the
information coming from the ({\it a priori} unknown) overall normalization.

\section{Accuracy and degeneracy}

How do these tests compare with regard to accuracy and degeneracy?
Their degeneracy structure is illustrated in Figure 3, which shows contour
plots of $\flum$, $\flum^2/E$, $t$, and $D/D(0)$ at three redshifts.
Using objects at a single redshift $z$, a test is unable to distinguish between
models lying along the same contour curve. The luminosity and size tests have
identical degeneracy structure because both probe $S(\kappa\I)/\kappa$;
their degeneracy curves are seen to rotate 
anti-clockwise from a slope of $1/2$ (explained below in \S\ref{q0Sec})
at $z=0$ to negative at $z=\infty$. 
The count contours rotate in the same sense as $z$ increases.
The isochrones rotate similarly but have a richer structure at $z=0$ because 
the age of the Universe probes $E$ at all redshifts.
They become vertical at high redshift where 
the age is independent of $\Ol$.
The growth factor degeneracy curves are seen to have
a slope steeper than $-1$ in most of our parameter space. 
This is because increasing the hyperbolic
curvature $1-\Om-\Ol$ makes fluctuation
growth freeze out earlier, increasing $D(z)/D(0)$, 
and increasing $\Ol$ at fixed curvature typically has the same effect.
The evolution of cluster abundance places powerful 
constraints on $D(z)/D(0)$ (Bahcall \& Fan 1998).
Although this test gives highly non-Gaussian
errors $\varepsilon_n$ (the constraints are mainly one-sided),
its degeneracy structure is still given by 
{\frenchspacing Fig.} 3.

The list of geometry tests that we have discussed is far from complete.
For instance, nonlinear effects in weak lensing (Jain \& Seljak 1997) and strong
lensing (Falco {\etal} 1998; Bartelmann {\etal} 1998) are promising probes of $\Om$ and $\Ol$. 
With CMB fixing other parameters, 
baryonic features detected in future galaxy redshift surveys
would give fairly vertical degeneracy curves, 
potentially measuring $\Omega_m$ to percent levels 
(Eisenstein {\etal} 1998).

For all tests modeled above, the {\it size} of the 
error ellipses scales as $\sigma/\sqrt{N}$, whereas the 
{\it shape} (slope and eccentricity) is given by the degeneracy 
structure.
The CMB ellipses in Figure 2 have been computed as in 
Eisenstein {\etal} (1998), marginalizing over 10 
additional
parameters.
This CMB information on 
$\Om$ and $\Ol$ comes mainly from the angular location of acoustic
features in the power spectrum, which depends principally on the 
curvature term $\kappa$, \ie, on the combination $\Om+\Ol$.

\centerline{{\vbox{\epsfxsize=9.0cm\epsfbox{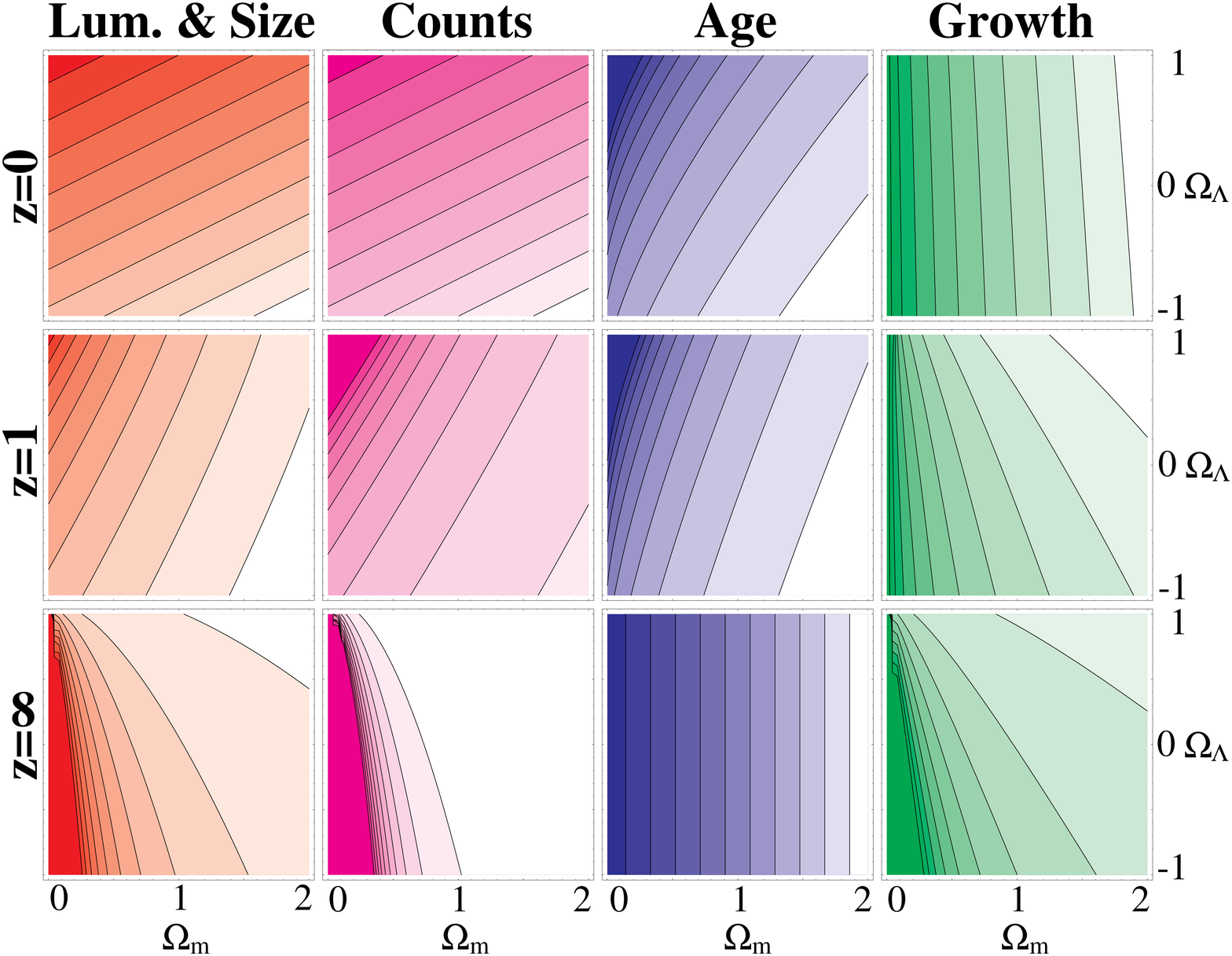}}}}
{\footnotesize {\bf FIG. 3}
--- How the degeneracy structure of different cosmological tests
rotates with redshift. All 12 panels have same axes.
}

\subsection{Low redshift observations such as SDSS}

\label{q0Sec}

It is well known that if data is available only for 
$z\ll 1$, then to first order, the luminosity, angle and count tests probe only
the parameter combination $q_0\equiv\Om/2-\Ol$.
In this limit, our results reduce to 
\beq{qFisherEq}
\F=
\left(\begin{tabular}{rc}
$1/4$&$-1/2$\\
$-1/2$&$\hphantom{-}1$
\end{tabular}\right)
(\Delta q_0)^{-2},
\eeq
where
$\Delta q_0=2\ln 10\Delta m/5N^{1/2}\zrms$ 
for luminosity tests using objects at {\rms} redshift of $\zrms$
with magnitude errors $\Delta m$,  
$\Delta q_0=2\sigma/N^{1/2}\zrms$ 
for corresponding angular size tests on objects with 
fractional size errors $\sigma$,
and 
$\Delta q_0=2(5/3\nbar)^{1/2}/\zmax$ 
for number count tests
volume limited to $\zmax$.
This is why the corresponding $z=0$ panels in Figure 3 both give the same
slope 1/2.

Because of this scaling, the huge number of galaxies in upcoming
surveys such as SDSS and 2dF may allow them to place competitive
constraints on $q_0$, as shown in Table 1, despite being a factor of
several below SN Ia in redshift.  Here we have assumed that fitting a
Schechter luminosity function to $N$ galaxies at the same redshift
determines the parameter $L_*$ to within $5/N^{1/2}$ magnitudes, which
is conservative based on Table 2 in Lin {\etal} (1996).  An obvious
obstacle to such measurements is that galaxy evolution (in luminosity,
size and number density) can mimic a change in $q_0$.  However, the
brute force statistical power of these data sets is so large that
even subsamples of 1\% of the galaxies give interesting constraints.
Studying how the ``$q_0$''-estimates vary as the galaxies are
subdivided by, \eg, morphology, luminosity and surface brightness
therefore holds the
potential of providing interesting information about galaxy
evolution and perhaps the true $q_0$.

\section{Conclusions}

In conclusion, we have derived useful expressions for 
the Fisher information matrix for a number of classical cosmological tests
and combined them with the Fisher matrix of the CMB.
Whereas two identical data sets give only a factor of $\sqrt{2}$ improvement in 
error bars when combined, the gain factor was found to exceed 10 when combining 
SN Ia with CMB.
This ``cosmic complementarity'' is due to the fortuitous fact that 
although either data set alone suffers from a 
serious degeneracy problem, the directions in which they 
are insensitive (in which the
ellipses in Figure~2 are elongated) are almost orthogonal.
The complementarity is even more dramatic for a standard
$\Om=1$, $\Ol=0$ CDM cosmology (Tegmark {\etal} 1998), where
a smaller ISW effect worsens the CMB degeneracy.

Figure 3 shows that this complementarity is rather generic, with 
degeneracy curves in virtually all directions.
This means that when three different tests are combined, 
there will be an important cosmic consistency check. If three skinny ellipses fail
to overlap, at least one measurement must be wrong, whereas if they
all cross at the same point, even hardened sceptics are likely to be impressed.

The potential power of upcoming CMB measurements has led to a widespread feeling that they
will completely dominate cosmological parameter estimation, with other types
of experiments making only marginal contributions.
Because of cosmic complementarity, of which the present paper gives 
a number of examples,
this view is misleading:
two data sets combined can be much more useful than either one alone.




\section*{Acknowledgments}
We thank David Hogg, Alex Kim, Robert Kirshner, Saul Perlmutter, 
Doug Richstone, Adam Riess and Martin White for 
useful discussions. 
MT was supported by NASA through grant NAG5-6034 and Hubble Fellowship
HF-01084.01-96A from STScI, which is operated by AURA, {\frenchspacing Inc.} 
under NASA contract NAS5-26555.
DJE is a Frank and Peggy Taplin Member at the IAS, and
WH is supported by the Keck Foundation and a Sloan Fellowship.
DJE and WH are also supported by NSF-9513835.


{
\footnotesize
Table 1 ---
Attainable error bars $\Delta\Omega_i$ for various combinations of data
sets.  The rows correspond to using CMB alone, three forecasts
(pessimistic, middle-of-the-road, and optimistic) for available SN Ia
data in five years time, and the SDSS tests described in the text.
The CMB columns correspond to the upcoming MAP and Planck satellite
missions without ($-$) and with $(+)$ polarization information.
Planck$+$ is seen to improve over the ``No CMB'' column by 
about an order of magnitude in $\Delta\Ol$,
and the difference is even greater between the
``Opt'' and ``No SN'' rows.  
The ``No SN'' row is overly conservative, since gravitational lensing 
breaks the CMB degeneracy somewhat 
(Metcalf \& Silk 1998; Stompor \& Efstathiou 1998)
but this lensing information is 
dwarfed by the SN Ia in the other rows. 
\begin{center}
\begin{tabular}{|l|cccc|cc|cc|cc|cc|cc|}
\hline
&&&&&\multicolumn{2}{c|}{No CMB}&\multicolumn{2}{c|}{MAP$-$}&\multicolumn{2}{c|}{MAP$+$}&\multicolumn{2}{c|}{Planck$-$}&\multicolumn{2}{c|}{Planck$+$}\\
Test		&$N$	&$\dm$	&$\zbar$&$\dz$	&$\dOm$	&$\dOl$	&$\dOm$	&$\dOl$	&$\dOm$	&$\dOl$	&$\dOm$	&$\dOl$	&$\dOm$	&$\dOl$	\\
\hline	
No SN Ia	&0	&$-$	&$-$	&$-$	&$\infty$&$\infty$&1.4	&1.1	&.25	&.20	&1.2	&.96	&.14	&.11	\\
Pess SN Ia	&100	&0.5	&0.55	&0.2	&.52	&.68	&.07	&.06	&.07	&.06	&.07	&.06	&.06	&.05	\\
Mid SN Ia	&200	&0.3	&0.65	&0.3	&.13	&.21	&.03	&.04	&.03	&.03	&.03	&.02	&.03	&.02	\\
Opt SN Ia	&400	&0.2	&0.70	&0.4	&.05	&.08	&.02	&.03	&.01	&.02	&.01	&.01	&.01	&.01	\\
\hline
SDSS $L_*$	&$10^6$	&5	&\multicolumn{2}{|c|}{$\zrms=0.1$}&$\infty$&$\infty$	&.04	&.02	&.03	&.02	&.01	&.01	&.005	&.004	\\
SDSS counts	&$5\times 10^5$&$-$&\multicolumn{2}{|c|}{$\zmax=0.1$}&$\infty$&$\infty$	&.05	&.03	&.02	&.01	&.03	&.02	&.03	&.02	\\
\hline
\end{tabular}
\vspace{-0.2cm}
\end{center}
}


\end{document}